\documentclass[prb,showpacs,twocolumn,floatfix]{revtex4}
\usepackage[english]{babel}
\usepackage{graphicx}
\usepackage{amsmath}
\usepackage{amssymb}

\newcommand{\Tr}{{\rm Tr}}

\begin{document}
\title{Dephasing-enabled triplet Andreev conductance}
\author{B. B{\'e}ri}
\affiliation{Instituut-Lorentz, Universiteit Leiden, P.O. Box 9506,
2300 RA Leiden, The Netherlands}
\date{March 2009}
\begin{abstract}
We study the conductance of  normal-superconducting quantum dots with strong spin-orbit scattering, 
coupled to a source reservoir using a single-mode spin-filtering quantum point
contact. The choice of the system is guided by the aim to study triplet
Andreev reflection without relying on half metallic materials with specific
interface properties. 
Focusing on the zero temperature, zero-bias regime, we show how dephasing due to
the presence of a voltage probe enables the conductance, which vanishes in the quantum limit, to take
nonzero values.
Concentrating on chaotic  quantum dots,  we obtain the full distribution
of the conductance as a function of the dephasing rate. 
As dephasing gradually lifts the conductance from zero, the dependence of the
conductance fluctuations on the dephasing rate is nonmonotonic.
This is in contrast to chaotic quantum dots in usual
transport situations, where dephasing monotonically suppresses the conductance fluctuations.
\end{abstract}
\maketitle{}

\section{Introduction}

\begin{figure}
\begin{center}
\includegraphics[height=4.5cm]{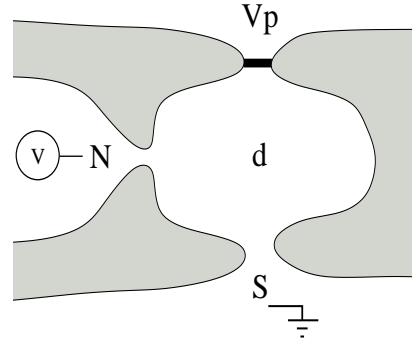}
\end{center}
\caption{Sketch of the setup studied in this paper. A normal conducting
  chaotic quantum dot (d) coupled to a superconductor (S) via a many-mode
  contact, and to a normal source reservoir (N), held at an infinitesimal voltage $V$, via a single-mode, spin-filtering
  QPC. Dephasing is introduced by coupling the dot to a voltage
  probe (Vp) via a contact supporting $N_\phi$ modes with a tunnel
  barrier (black rectangle) of transparency $\Gamma_\phi$ per mode.}
\label{fig:setup}
\end{figure}

The triplet superconducting proximity effect\cite{Berg01a,Kadig01,bergeretRMP}
in half-metals (fully polarized ferromagnets, conducting only for one spin direction)
has received a  considerable  attention recently, both 
theoretically\cite{eschrig2003thm,Braude07,asano07,asano07PRB,takahashi07,eschrig2008tsc,Galak08,BKBB09}
and  experimentally.\cite{keizer2006sts,Krivor06epl,Yates07,Krivor07}
The mechanism behind the effect is the process of triplet Andreev  reflection
at the   half-metal--superconductor interface.\cite{Kadig01,eschrig2003thm} 
The key ingredient that allows and influences this reflection process is
provided by the  magnetic properties
of the interface between the half-metal and the superconductor: it should have a  magnetization that is 
misaligned from that of the half-metal.\cite{Kadig01,eschrig2003thm}
The role of such an interface is to break all the symmetries in spin-space, thereby allowing
for the spin rotations necessary for the triplet Andreev reflection.
The properties of the interface, however, are not easy to manipulate in experiments,
which is the reason why only a low proportion of half-metal--superconductor
samples show behavior consistent with triplet Andreev reflection.\cite{Krivor06epl,Yates07,Krivor07} 
Here we study  triplet Andreev reflection in a setup
that is free of this difficulty.  The setup consists of  an Andreev quantum
dot\cite{Bee05} (i.e., a quantum dot in contact with a superconductor), coupled
to a normal, source reservoir via a single-mode quantum point contact (QPC), see
Fig.~\ref{fig:setup}.
Spin-orbit scattering in the quantum dot is assumed to be strong enough
that the direction of the spin is randomized in much shorter time than the
typical time $t_{\rm dw}$ of the escape from the dot. This
allows the dot to effectively play the role of the interface.
The role of the half metal is  played by the QPC, which is set to the spin-selective $e^2/h$ conductance
plateau using a parallel magnetic field.\cite{Potok02}
(For simplicity, we assume that the Andreev conductance of the contact to the
superconductor is much larger than $e^2/h$, which makes the transport
properties insensitive to the details of this contact.)

A surprising feature of triplet Andreev reflection is that despite the
randomized spin in the quantum dot, the conductance of
such a fully phase coherent, single-channel system vanishes  in  the zero
temperature, linear response regime.\cite{BKBB09}
While  current can be passed through the system  using finite voltages
or temperature, it is natural to ask, whether there is still a possibility  
for transport in the zero temperature, linear response limit.
In this paper we show that there is: relaxing the condition of full phase
coherence enables the zero-bias triplet Andreev conductance to take nonzero
values. In the remaining sections,  our goal is to demonstrate this statement by
studying the behavior of the triplet Andreev conductance in the presence of
dephasing in detail.

\section{Voltage probe as a source of dephasing}

We introduce dephasing by coupling the quantum dot to an additional normal
reservoir, which acts as a voltage
probe.\cite{Buett86a,Buett88a,marcus93,BarMel95,BroBee97volprobe}
A voltage probe draws no net current, but it absorbs and
reinjects quasiparticles without a phase relationship, thereby destroying
phase coherence.
Dephasing due to a voltage probe in normal-superconducting structures at zero
temperature was studied in Refs.~\onlinecite{mortensen2000dss,gramespacher2000dfa,Bel03}, for
systems where no triplet Andreev reflection could occur.
The contact to the voltage probe is characterized by the number of modes  (including the spin degrees of freedom) 
$N_\phi$ and the tunnel probability per mode  $\Gamma_\phi$, 
which together determine the dephasing rate as
$g_\phi=N_\phi\Gamma_\phi\delta/h$, where 
$\delta$ denotes the mean level spacing of the quantum dot.
We consider two limits, a voltage probe with a single mode,
spin-filtering contact, $N_\phi=1$, and a voltage probe with macroscopic number of
modes, $N_\phi\gg 1$. In the first case the dephasing rate is controlled by
the tunnel probability. 
In the second case, for $g_\phi\sim 1$, which
will turn out to be the  regime where the conductance behaves nontrivially, the transport properties
depend on $N_\phi$ and $\Gamma_\phi$ only through their product, i.e.,  the dephasing rate.\cite{BroBee97volprobe} 
The two limits considered here represent two types of voltage
probes that appear in the context of spin dependent quantum transport.\cite{Bee06excess}
The probe $N_\phi=1$ is a  spin-conserving probe, while without further
specification, the $N_\phi\gg 1$ can be either a spin-conserving or a
spin-nonconserving voltage probe. For the systems studied in
this paper, there is no need for further specification, as the type
of the voltage probe  is unimportant due to the strong spin-orbit
scattering in the quantum dot. 
 
We formulate our problem within the framework of the scattering matrix approach. 
The transport quantities of interest are expressed in terms of the scattering
matrix $S$  at the Fermi energy (the chemical potential of the
superconductor), relating incoming and outgoing modes (including the electron-hole
degrees of freedom) in the contacts to the normal reservoirs. 
The currents at the contact to the  source ($s$) and the voltage probe
($\phi$) are given by\cite{takane1992cfm,lambert1993mpc}
\begin{subequations}
\begin{align}
I_\alpha&=\frac{e^2}{h}\sum_{\alpha\beta}\left[N_\alpha\delta_{\alpha\beta}
+{\cal R}^{ he}_{\alpha\beta}-{\cal R}^{ ee}_{\alpha\beta}\right]V_\beta\\
{\cal R}^{ij}_{\alpha\beta}&=\Tr\left[(S^{ ij}_{\alpha\beta})^\dagger S^{ ij}_{\alpha\beta}\right],
\end{align}
\end{subequations}
where $\alpha, \beta=s,\phi$, and the index $e, h$
refers to electron and hole modes, respectively. 
The voltages $V_\beta$ are measured from the chemical potential of the
superconductor, which is assumed to be grounded.
The voltage
$V_\phi$ is determined by demanding that no  current is drawn to the
voltage probe, $I_\phi=0$. The conductance, defined by $G=V_s/I_s$ is given by
\begin{equation}
\frac{h}{e^2}G=1+{\cal R}^{he}_{ss}
-{\cal R}^{ee}_{ss}- \frac{({\cal R}^{he}_{s\phi}  -{\cal R}^{ee}_{s\phi})
      ({\cal R}^{he}_{\phi s}-{\cal R}^{ee}_{\phi s})}
{N_\phi+{\cal R}^{he}_{\phi \phi} -{\cal R}^{ee}_{\phi \phi}},
\label{eq:condgen}
\end{equation}
where we substituted $N_s=1$. Equation \eqref{eq:condgen} is the starting
point for our calculations. In what follows, we concentrate on systems where the motion inside
the quantum dot is chaotic. We are interested in the statistics of the conductance, which 
we obtain using Random Matrix Theory\cite{RMTQTR} for the scattering matrix
$S$.

\section{Dephasing due to a single mode voltage probe}

We first discuss the case voltage probe with $N_\phi=1$. 
The calculational advantage of this case is that it allows for  a problem with
minimal dimension, with the single mode source contact and a single mode
voltage probe contact resulting in a $4\times 4$ scattering matrix. 
The parallel magnetic field together with the strong spin-orbit scattering 
places the quantum dot in the unitary symmetry class.\cite{AF01} Consequently, 
the dot-superconductor system belongs to class $D$ in the symmetry
classification of Altland and Zirnbauer,\cite{Alt97} which translates to $S=\Sigma_1 S^*
\Sigma_1$ as the only constraint for the scattering matrix, besides unitarity.
($\Sigma_j$ denotes the $j$-th Pauli matrix in electron-hole space.)
Assuming that the contact to the source reservoir is ideal, the  two single-mode QPC-s
can be characterized by the reflection matrix
\begin{equation}
r=\begin{pmatrix}
 0 &  0\\
 0 &         \sqrt{1-\Gamma_\phi} e^{i\Sigma_3 \xi},
\end{pmatrix}
\end{equation}
where the block structure reflects a grading according to the normal contacts and 
$\xi$ is the reflection phase shift for electrons at
the voltage probe contact.
The statistical properties of the conductance follow from the distribution of
the scattering matrix, which is given by the generalization of the Poisson
kernel,\cite{BBPoisson}
\begin{equation}
P(S)\propto |\det(1-r^\dagger S)|^{-3}.
\label{eq:Poisker}\end{equation}
The probability distribution is understood with respect to the invariant
measure $d\mu(S)$ on the manifold ${\cal M}_D$ defined by  $S=\Sigma_1 S^*
\Sigma_1$ in the space of $4\times4$ matrices.
We parametrize $S$ as
\begin{equation}
S=\begin{pmatrix}
e^{i\psi_1}\sqrt{1-T}\openone_2  & e^{i\psi_2}\sqrt{T}\tau\\
e^{-i\psi_2}\sqrt{T}\tau& e^{-i\psi_1}\sqrt{1-T}\openone_2 
\end{pmatrix}
\begin{pmatrix}
W & 0\\
0 & W^*
\end{pmatrix},
\label{eq:Sparam}
\end{equation}
where $T\in (0,1)$, $\psi_1,\psi_2\in(0,2\pi)$, $W\in$ SU$(2)$ and
$\tau=i\sigma_2$. ($\sigma_j$ denotes the $j$-the Pauli matrix in spin-space.)
The matrix structure in \eqref{eq:Sparam} corresponds to electron-hole grading.
The above parametrization can be obtained from the polar decomposition
introduced in the Appendix. Eq.~\eqref{eq:Sparam} implies that $\det(S)=1$ and
that the matrix $S^{he}(S^{he})^\dagger$ has a twofold degenerate eigenvalue
$T$.  This is true for the generic setups with vanishing linear
conductance in the fully phase coherent limit, i.e., if the closed Andreev
quantum dot has no energy level at the Fermi energy.\cite{BKBB09} (For a
detailed discussion of this point we refer to the Appendix.)
Using the Euler angle parametrization for $W$, 
\begin{equation}\begin{array}{c}
W\!\!=\!\!\!\!
\ 
\left(\begin{array}{cc}
e^{-i(\varphi+\psi)/2}\cos(\theta/2) &-e^{i(\psi-\varphi)/2}\sin(\theta/2)\\
e^{i(\varphi-\psi)/2}\sin(\theta/2)&e^{i(\varphi+\psi)/2}\cos(\theta/2)
\end{array}\right)\!\!,\\
\quad \\
(\varphi,\psi,\theta) \in [0,2\pi]\times [0,4 \pi]\times [0,\pi],
\end{array}\end{equation}
the invariant measure on ${\cal M}_D$ is given by $d\mu(S)\propto
\sin(\theta)$, and the conductance in units of $e^2/h$ is
\begin{equation}
G(T,\theta)=4\left(\frac{1}{T}+\frac{1}{\sin^2(\theta/2)}\right)^{-1}.
\end{equation}
The distribution of the conductance is given by \mbox{$P_{\Gamma_\phi}(G)=\int
  d\mu(S)P(S)\delta(G-G(T,\theta))$}, which can be reduced to
\begin{equation}
\begin{array}{c}
\displaystyle P_{\Gamma_\phi}(G)=\frac{\Gamma_\phi^3}{2
  G^2}\int_1^{4/G-1}\!\!\!\!\!\!\!\!\!\! dx \frac{2a^2+b^2}{x^2(4/G-x)^2(a^2-b^2)^{5/2}},\\
\quad \\
\displaystyle a=1-(1-\Gamma_\phi)\left(\frac{1}{4/G-x}+\frac{1}{x}-1\right),\\
\quad \\
\displaystyle \frac{b^2}{4}=(1-\Gamma_\phi)\left(1-\frac{1}{4/G-x}\right)\left(1-\frac{1}{x}\right)
\end{array}
\label{eq:Gphiint}\end{equation}
for $0\leq G \leq 2$ and $0$ otherwise. A closed form expression can be given for $\Gamma_\phi=1$, 
\begin{figure}
\begin{center}
\includegraphics[height=5.5cm]{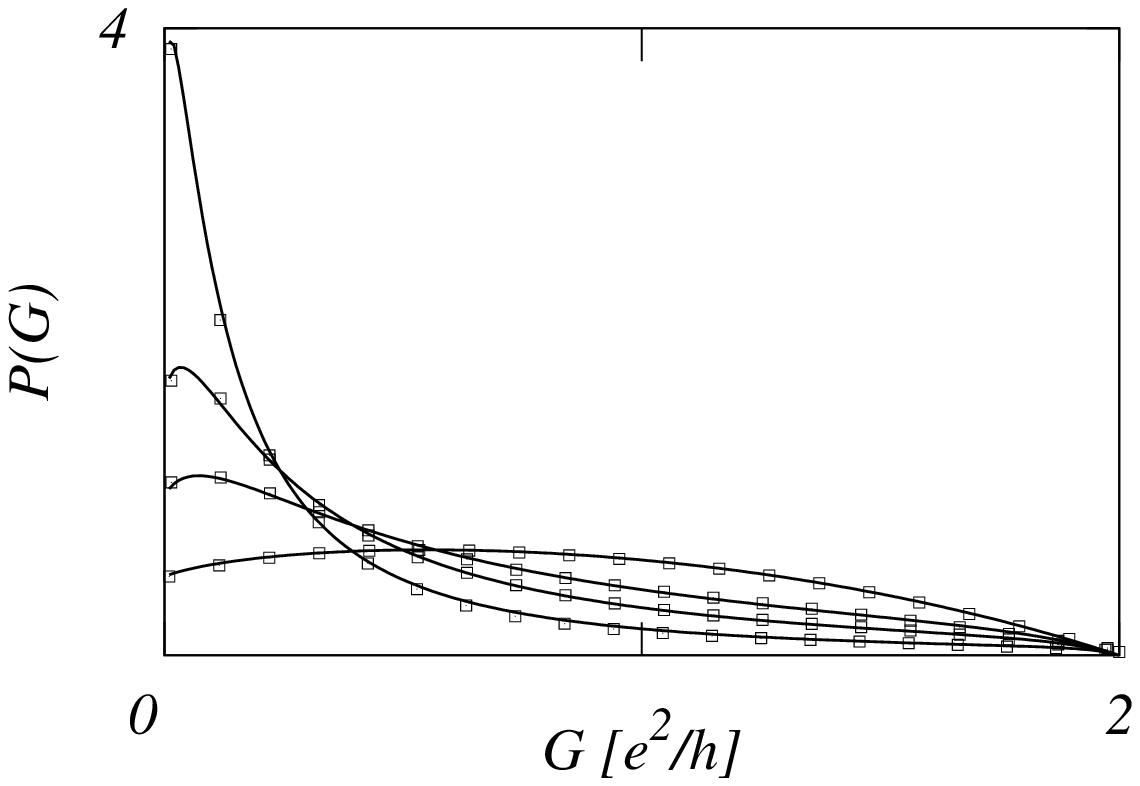}
\includegraphics[height=5.5cm]{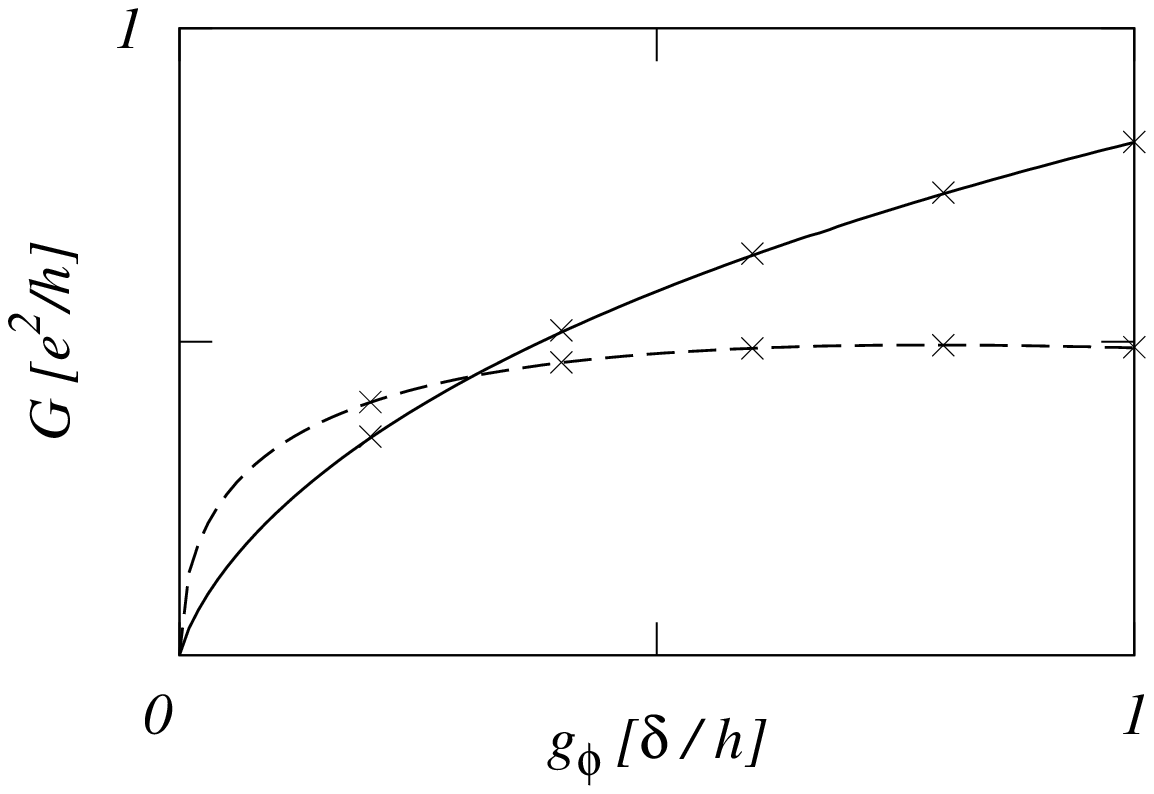}
\end{center}
\caption{Upper panel: Probability distribution [Eq.~\eqref{eq:Gphiint}] of the conductance for
  $N_\phi=1$, for various values of the  dephasing rate $g_\phi=\Gamma_\phi\delta/h$. 
        The curves, in order of decreasing maximum, correspond to $\Gamma_\phi=0.2$, $0.4$, $0.6$, $1$, respectively.
        The empty squares represent smoothed histograms obtained from 3000 numerically
        generated scattering matrices  for each value of $\Gamma_\phi$, for a
        system where the  superconducting contact supports 50
        propagating modes.
        Lower panel: the average  (solid line) and the
        standard deviation  (dashed line) of the conductance as a function of $g_\phi$.
        The crosses are results of the numerical simulation.}
\label{fig:contrdeph}
\end{figure}

\begin{equation}
P_{\Gamma_\phi=1}(G)=1-\frac{2}{4-G}-\frac{G}{4}\ln \frac{G}{4-G}.\label{eq:distroG1}\end{equation}
For $0<\Gamma_\phi<1$, we evaluated the integral \eqref{eq:Gphiint}
numerically. The resulting distribution is shown in the top panel of
Fig.~\ref{fig:contrdeph} for several values of $\Gamma_\phi$. 
In the absence of dephasing, the conductance vanishes, corresponding to
$P_{\Gamma_\phi=0}(G)=\delta(G)$. 
With the gradual introduction of dephasing, $G$ is  enabled to take nonzero
values, leading to a widening of the conductance distribution with
increasing dephasing rate, eventually reaching the distribution
\eqref{eq:distroG1} for $\Gamma_\phi=1$. 
In the bottom panel of Fig.~\ref{fig:contrdeph} we show the dependence of
the average and the variance of the conductance on
$\Gamma_\phi$. 
While the average conductance increases monotonically with increasing
dephasing rate, the variance increases to a maximum at $\Gamma_\phi\approx 0.8$, which
is followed by a slight decrease. 
The finite value of the conductance fluctuations at $\Gamma_\phi=1$
(corresponding to the maximal dephasing for $N_\phi=1$) indicates
that a single channel voltage probe can not
lead to a complete loss of phase coherence -- as we will see
below, without phase coherence, the  conductance fluctuations are suppressed back to zero.
For weak dephasing, $\Gamma_\phi\ll 1$, the conductance
distribution is rapidly decaying away from $G=0$ and it has the scaling form
$P_{\Gamma_\phi}(G)= f(G/\Gamma_\phi)/G$. This results in the dependence
\begin{equation}
  \langle G^n \rangle\propto \Gamma_\phi^n,\quad \Gamma_\phi\ll 1
\end{equation}
for the $n$-the moment of the conductance.

\section{Dephasing due to a multimode voltage probe}

\begin{figure}
\begin{center}
\includegraphics[height=5.5cm]{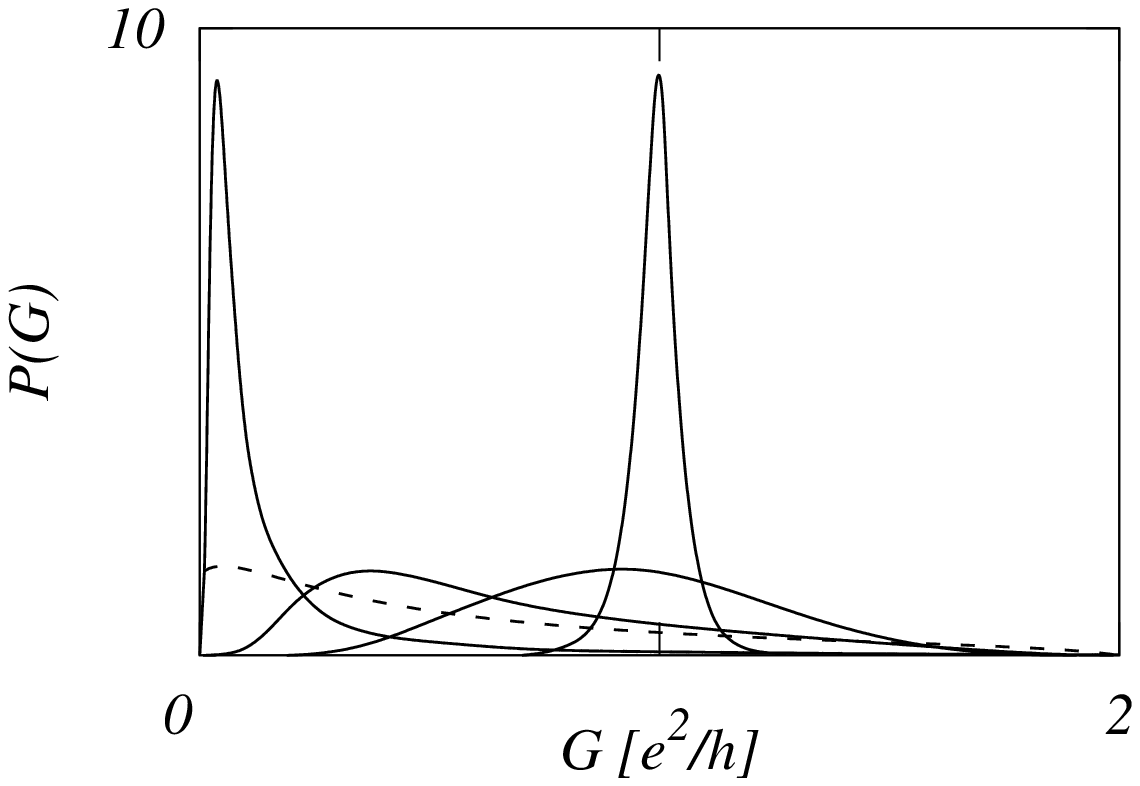}
\includegraphics[height=5.5cm]{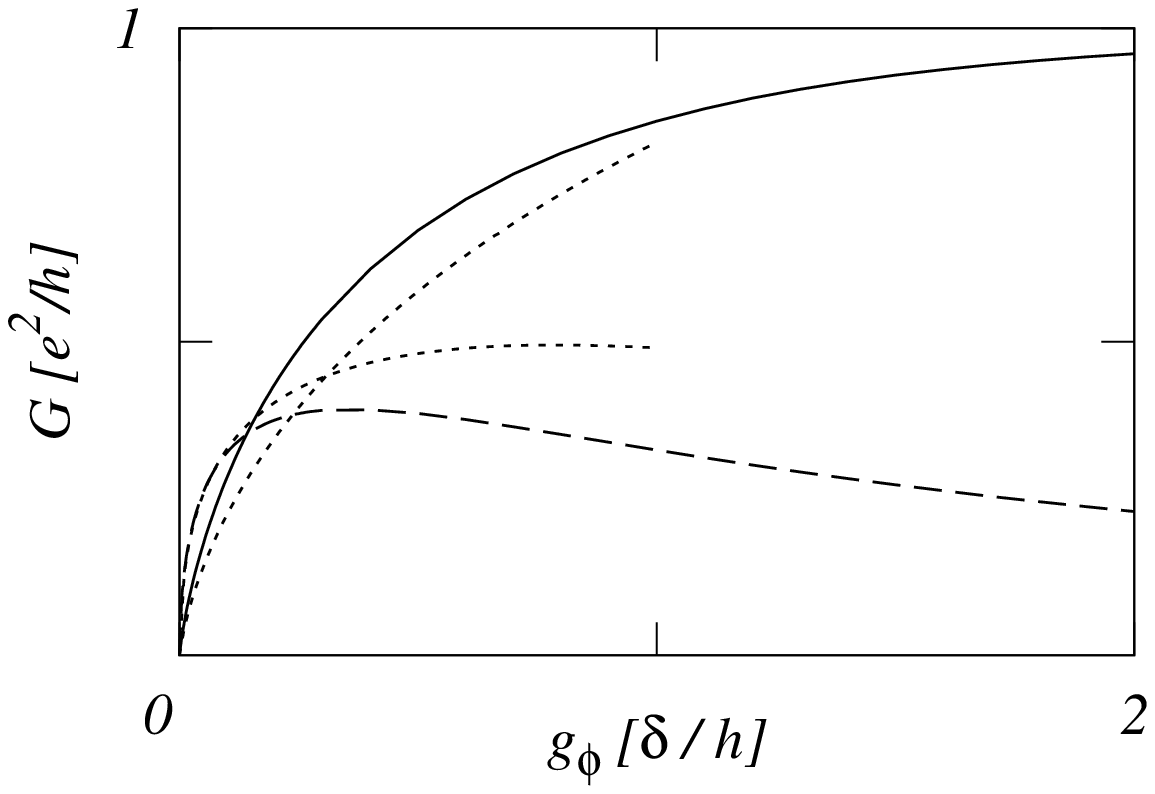}
\end{center}
\caption{Upper panel: Probability distribution of the conductance for
  $N_\phi\gg 1$, for different values of  the dephasing rate $g_\phi$. 
        The solid curves, in order of increasing position $G$ of the maximum,
        correspond to  $g_\phi h/\delta=0.05$, $0.5$, $1.5$, $10$, respectively.
        The curves are obtained by smoothing  histograms from 3000 numerically
        generated  scattering matrices for each value of $g_\phi$, with
        $N_\phi=100$, for a system where the superconducting contact supports 50
        propagating modes.
        For comparison, we show the distribution for $N_\phi=1$, $g_\phi h/\delta=0.5$ (dashed line).
        Lower panel: the average  (solid line) and the
        standard deviation  (dashed line) of the conductance as a function of $g_\phi$.
        For comparison, the dotted lines show the corresponding functions for $N_\phi=1$.}
\label{fig:intrdeph}
\end{figure}

Now we turn to  the case of the voltage probe with macroscopic number of
channels, $N_\phi\gg 1$. While it might be possible to make some analytical progress using
the Poisson kernel distribution of Ref.~\onlinecite{BBPoisson} and following similar steps to the
calculation of Ref.~\onlinecite{BroBee97volprobe}, we resort to a simpler approach and
obtain the  statistics of the conductance by generating an ensemble of scattering matrices
 numerically. The scattering matrix $S$ is expressed in terms of the electron scattering matrix
\begin{equation}
S_N=
\begin{pmatrix}
r & t'\\
t & r'
\end{pmatrix}
\end{equation} 
of the normal region. Here $r$ is describes reflection from the dot through
the normal contacts, $r'$ describes reflection through the superconducting
contacts, $t$ corresponds to transmission to the superconducting, and $t'$ to
the normal contacts. 
The necessary blocks of $S$ in electron-hole grading are given
by\cite{Bee92a,RMTQTR}
\begin{subequations}
  \label{eq:reereh}
\begin{align}
S^{ee}&=r- t' \sigma_2 r'^*\sigma_2(1+ r' \sigma_2 r'^* \sigma_2)^{-1}t\ ,\\
S^{he}&=- t'^*\sigma_2(1+ r'\sigma_2 r'^*\sigma_2)^{-1}t\ ,\label{eq:rhe}
\end{align}
\end{subequations}
The scattering matrix $S_N$ can be expressed using the statistical mapping\cite{mello1985ita,Brouw95gce}
\begin{equation}
  S_N = \sqrt{1 - \Gamma} - \sqrt{\Gamma} {1 \over 1 - S_0
  \sqrt{1 - \Gamma}} S_0 \sqrt{\Gamma}, \label{eq:tunS}
\end{equation}
where $S_0$ is unitary and $\Gamma$ is a diagonal matrix containing the transmission probabilities
of the contacts with $\Gamma_{11} = 1$ corresponding to  perfect transmission through the single mode QPC
and $\Gamma_{jj} = \Gamma_\phi$ for $1<j\leq N_\phi+1$ describing tunneling at
the voltage probe. 
We took $\Gamma_{jj}=1$ for $j> N_\phi+1$, corresponding to the contact to the
superconductor. The results do not depend on this choice, as long as the
Andreev conductance of the contact is much greater than $e^2/h$.
Using the mapping \eqref{eq:tunS}, the  distribution of $S_N$ is
obtained by taking $S_0$ from the circular
unitary ensemble,\cite{mello1985ita,Brouw95gce} which we generated numerically.\cite{mezzadri2007grm}
For a mutual test of the  program  and the calculations, 
we first show results for $N_\phi=1$  in Fig.~\ref{fig:contrdeph}.
As it is seen, the agreement between the calculation and the numerics is
perfect. 
The conductance distribution in the limit $N_\phi\gg 1$  is shown 
in the top panel of Fig.~\ref{fig:intrdeph} for several values of the dephasing
rate $g_\phi$. The distribution $P_{g_\phi}(G)$ initially widens from
$P_{g_\phi=0}=\delta(G)$ with increasing $g_\phi$ and then it gradually
narrows again to $P_{g_\phi=\infty}=\delta(G-G_{\rm class})$, where
\begin{equation}
G_{\rm class}=\left(\frac{1}{G_{\rm QPC}}+\frac{1}{G_{\rm     NS}}\right)^{-1}\approx G_{\rm QPC}=1
\end{equation}
is conductance of the single mode QPC and the Andreev conductance of the 
superconducting contact in series, in units of $e^2/h$. 
The dependence of the average and the variance of the
conductance on $g_\phi$ is shown in bottom panel of
Fig.~\ref{fig:intrdeph}. While the average conductance increases monotonically
to its classical value, the conductance fluctuations display nonmonotonic
behavior, corresponding to the initial widening and
the final re-narrowing of the conductance distribution. 
Fig.~\ref{fig:intrdeph} also shows a comparison between the limits
$N_\phi=1$ and $N_\phi\gg 1$.  
For a given value of $g_\phi$, the conductance distribution close to
$G=0$ is suppressed for $N_\phi\gg 1$, in contrast to the single channel case,
where $P(G=0)$ is finite. 
The average conductance increases faster for $N_\phi\gg 1$ towards its
classical value, while the conductance fluctuations are suppressed compared to 
$N_\phi= 1$.

\section{Conclusions}
In summary, we have studied in detail how dephasing due to a
voltage probe enables a nonzero value for the  zero temperature, zero-bias
triplet Andreev conductance in Andreev quantum dots with a
single-channel spin-filtering source point contact. 
We focused on systems where the quantum dot is chaotic, and obtained the full 
distribution of the conductance as a function of the dephasing rate for two
limiting cases for the number of modes $N_\phi$ in the voltage probe contact, $N_\phi=1$, and $N_\phi\gg 1$.
Compared to chaotic quantum dots in other transport situations, our findings
for the conductance are quite unusual. 
Dephasing is known to monotonically suppress the conductance fluctuations, in general.\cite{BarMel95,BroBee95volprobe,Huib98dcpd,RMTQTR} 
In contrast, as dephasing gradually enables transport, the fluctuations of the
triplet Andreev conductance are initially enhanced, which is
followed by a suppression for strong dephasing, i.e., the overall dependence
on the dephasing rate is nonmonotonic.

It is worthwhile to point out that in the $N_\phi\gg 1$ case, unlike
Ref.~\onlinecite{BroBee95volprobe}, we did not
intend to use the voltage probe to model dephasing processes intrinsic to the quantum dot,
since accounting for the temperature dependence of such
processes would necessitate considering the effect of thermal smearing.\cite{Huib98dcpd}
Instead, our results apply to the situation where the dephasing rate is
controlled by  a real, physically present voltage probe. 
Experimental control of the dephasing rate using a voltage probe was demonstrated
very recently in the work of Roulleau {\it et al}.\cite{Roulleau} This makes us believe
that, in principle, it is realistic to test our predictions in experiments.

\acknowledgments
This work originated from discussions with P.~W.~Brouwer. The author has also
benefited from discussions with C.~W.~J.~Beenakker. This research was
supported by the Dutch Science Foundation NWO/FOM.

\appendix
\section{Electron-hole symmetry, polar decomposition and Andreev reflection eigenvalues}
The Andreev reflection eigenvalues $T_j$ are the eigenvalues of the matrix
$S^{he}(S^{he})^\dagger$. We prove here the consequences of  electron-hole symmetry on
these quantities, and relate them to the condition of the absence of energy level of the closed 
Andreev quantum dot at the Fermi energy.
\ \\

\noindent{\bf Theorem:}
At the Fermi energy,  the degeneracy $d_{j}$ of the Andreev
reflection eigenvalue $T_j$  is even if \mbox{$T_j(1-T_j)\neq 0$}, and 
$\det(S)=(-1)^{d_{u}}$, where $d_{u}$ is the degeneracy of the unit Andreev
reflection eigenvalue, if present, $d_u=0$ otherwise. 
Furthermore, the scattering matrix at the Fermi energy can be decomposed in electron-hole grading as 
\begin{equation}
S=
\begin{pmatrix}
U& 0\\
0 & U^*
\end{pmatrix}
\begin{pmatrix}
\hat R & \hat{\rho}\hat{T}\\
\hat{\rho}\hat{T}& \hat{R}
\end{pmatrix}
\begin{pmatrix}
V& 0\\
0 & V^*
\end{pmatrix},
\label{eq:poldec}
\end{equation}
where $U$ and $V$ are unitary matrices,
\begin{subequations}
\begin{align}
\hat{R}&=\bigoplus_j \sqrt{1-T_j}\openone_{d_{j}}\\
\hat{T}&=\bigoplus_j \sqrt{T_j}\openone_{d_{j}}, 
\end{align}
\end{subequations} 
and $\hat{\rho}=\bigoplus_j\rho_j$, where
\begin{equation}
\rho_j=\left\{ 
\begin{array}{ll}
\openone_{d_{j}} & \textrm{if } T_j(1-T_j)=0\\
\openone_{d_{j}/2}\otimes\tau &\textrm{otherwise.} \\
\end{array}
\right.
\end{equation}
\ \\ 

\noindent{\it Proof:} Following from the electron-hole symmetry \mbox{$S=\Sigma_1 S^* \Sigma_1$}, the
scattering matrix has the block decomposition
\begin{equation}
S=\begin{pmatrix}
S^{ee} & (S^{he})^*\\
S^{he} & (S^{ee})^*
\end{pmatrix}.
\label{eq:ehstruct}
\end{equation}
We start with the singular value decomposition 
\begin{equation}
S^{ee}=U'\hat{R}V',
\label{eq:Seedec}
\end{equation}
where  $U', V'$ are unitary
matrices. Using $(S^\dagger S)^{ee}=\openone$ and  $(S S^\dagger)^{ee}=\openone$, one finds that
\begin{equation}
S^{he}=U'^*Z\hat{T}V'.
\label{eq:Shedec}
\end{equation}
Here $Z$ is a block diagonal unitary matrix, 
\begin{equation}
Z=\bigoplus_j Z_j,\quad {\rm dim}Z_j=d_{j}.
\end{equation}
Substituting \eqref{eq:Seedec} and \eqref{eq:Shedec} into $(S^\dagger S)^{he}=0$  leads to 
\begin{equation}
\sqrt{T_j(1-T_j)}Z_j=-\sqrt{T_j(1-T_j)}Z_j^T
\label{eq:asy}
\end{equation}
For $T_j(1-T_j)\neq 0$, $Z_j$ is antisymmetric, and due to its unitarity
$\det (Z_j)\neq 0$, from which it follows that $d_{j}$ is even. Being
antisymmetric and unitary, $Z_j$ can be decomposed as\cite{youla1961nfm,Schliemann}
\begin{equation}
Z_j=U_j^T \hat{\tau} U_j,\quad \hat{\tau}=\openone_{d_{j}/2}\otimes \tau,
\end{equation}
where $U_j$ is unitary.
For $T_j=0,1$, Eqn.~\eqref{eq:asy} is automatically satisfied, without further requirements
for $Z_j$. For the zero Andreev reflection eigenvalue, if present, we can set $Z_{j}=U_{j}^TU_{j}$
with an arbitrary unitary matrix $U_{j}$, as for $T_j=0$, $Z_{j}$ drops out from \eqref{eq:Shedec}. 
For the unit Andreev reflection eigenvalue, if present, we write $Z_{j}=U_{j}^TU_{j}'$ with $U_{j}$, $U_{j}'$
unitary. Taken together, the matrix $Z$ can be written as
\begin{equation}
Z=\bigoplus_j U_j^T\rho_j U_j',
\end{equation}
where $U_j'=U_j$ for $T_j\neq 1$. Writing $U'$ and $V'$ as
\begin{subequations}
\begin{align}
U'&=U\ \left(\bigoplus_j U_j\right)\\
V'&=\left(\bigoplus_{j}U_j'^\dagger\right)\ V,
\end{align}
\end{subequations}
with $U$, $V$ unitary, one finds
\begin{subequations}
\begin{align}
S^{ee}&=U\hat{R}V\\
S^{he}&=U^*\hat{\rho} \hat{T}V,
\end{align}
\end{subequations}
which gives the decomposition \eqref{eq:poldec} upon substitution in Eq.~\eqref{eq:ehstruct}.
The decomposition \eqref{eq:poldec} satisfies the unitarity and electron-hole
symmetry requirements, therefore there are no further relations between the
matrices $U$ and $V$. The result $\det(S)=(-1)^{d_{u}}$ follows
straightforwardly. $\Box$ 

Note that in Eq.~\eqref{eq:Sparam}, we assumed $\det(S)=1$, however only
$\det(S)=\pm 1$ follows from $S=\Sigma_1 S^* \Sigma_1$. 
We show below that this is a valid assumption in the generic situation that
there is no energy level of the closed Andreev quantum dot  at the Fermi energy. 
Using the channel coupled model employed in  Ref.~\onlinecite{Alt97}, the
scattering matrix can be expressed as\cite{RMTQTR}
\begin{equation}
S_E=\frac{1+i\tilde{H}_E}{1-i\tilde{H}_E}.
\end{equation}
Here the Hermitian matrix $\tilde{H}_E=-\Sigma_1\tilde{H}_{-E}^*\Sigma_1$ 
is a projection of  $({\cal H}-E)^{-1}$, where
${\cal H}$ models the closed Andreev quantum dot. If ${\cal H}$ has no zero eigenvalues, i.e., there is no level at the
Fermi energy, the matrix $\tilde{H}_{E}$ can be taken at $E=0$ without
complications. Following from the symmetry of $\tilde{H}_{E=0}$, the
eigenvalues of $S$ come in complex conjugate pairs,  therefore, $\det(S)=1$.
(If there is a level at the Fermi energy, an eigenvalue of $\tilde{H}_{E}$ can
diverge as $E\rightarrow 0$ while an other can tend to zero,  leading to
a $(1,-1)$ eigenvalue pair of $S$, and thereby to $\det(S)=-1$.)
This result, together with $\det(S)=(-1)^{d_{u}}$, contains as a special case
the finding of Ref.~\onlinecite{BKBB09}, that for a single mode system, 
$S^{he}=0$ at the Fermi level, if the closed Andreev quantum dot  has no
level at the Fermi energy. Indeed, for such a system, $S$ is a $2\times 2$
matrix, i.e.,  there is a single Andreev reflection eigenvalue. 
As it is singly degenerate, it can be only zero or unity, and $\det(S)=1$
guarantees that it is zero.  For the $4\times 4$ matrix in Eq.~\eqref{eq:Sparam}, 
the degeneracy of the Andreev reflection eigenvalue also follows from
$\det(S)=1$. If there was no degeneracy, the eigenvalues could be only a zero and
a unit eigenvalue. This would mean  $\det(S)=-1$.

\end{document}